\documentclass[twocolumn,aps,prd,nofootinbib]{revtex4-1}

\pdfoutput=1

\usepackage{amsmath, amsfonts,amsthm, amssymb, mathrsfs, bm, verbatim, pslatex, bigints, physics, tikz, xcolor, graphicx,tikz-feynman}
\usetikzlibrary{decorations.markings}

\newcommand{\be}{\begin{equation}}
\newcommand{\ee}{\end{equation}}
\newcommand{\bea}{\begin{eqnarray}}
\newcommand{\eea}{\end{eqnarray}}
\newcommand{\K}{M}
\newcommand{\Kr}{J}
\newcommand{\lamC}{\lambda_C}
\newcommand{\vac}{\mbox{vac}}
\newcommand{\Aphi}{A_{\varphi}}
\newcommand{\cphi}{\phi}
\newcommand{\phic}{\theta_c}
\newcommand{\phiz}{\theta_0}
\newcommand{\uv}{\epsilon}

\begin{document}

\title{Counting of States in Higgs Theories}

\author{Mark P. Hertzberg}
\email{mark.hertzberg@tufts.edu}
\affiliation{Institute of Cosmology, Department of Physics and Astronomy,
Tufts University, Medford, MA 02155, USA}

\author{Mudit Jain}
\email{mudit.jain@tufts.edu}
\affiliation{Institute of Cosmology, Department of Physics and Astronomy,
Tufts University, Medford, MA 02155, USA}

\begin{abstract}
We enumerate the micro-states in Higgs theories, addressing (i) the number of vacuum states and (ii) the appropriate measure in the quantum path integral. To address (i) we explicitly construct the set of ground state wave-functionals in the field basis focussing on scalar modes $\theta(x)$. Firstly, we show that in the limit in which the gauge coupling is zero, we obtain an infinite set of degenerate ground states at large volume distinguished by $\theta(x)\to\theta(x)+\theta_0$, spontaneously breaking the global symmetry, as is well known. We then show that at finite gauge coupling there is a unique ground state at large volume since the wave-functional only depends on $\nabla\theta$ in the IR, and we explain this at the level of the Lagrangian. Since gauge fields fall off exponentially from sources there are no conserved charges or symmetries in the Higgs phase; so the Higgs mechanism is the removal of symmetry from the theory. We show how physical features of defects, such as cosmic strings in the abelian Higgs model, are best understood in this context. Since there is a unique ground state, we address (ii) whether the path integral is a volume measure for the radial Higgs field $\mathcal{D}\rho\,\rho^{N-1}$ from the $N$ components of the Higgs multiplet, or a line measure $\mathcal{D}\rho$ as the $N-1$ would-be Goldstones can be removed in unitary gauge. We prove that the need to avoid quartic divergences demands a tower of counter terms that resum to exactly give the volume measure. So the size of the Hilbert space in the zero gauge coupling case and finite gauge coupling case are in one-to-one correspondence, despite the degeneracy of the ground state being lifted in the latter. As a cosmological application, we point out that the volume measure can make it exponentially more unlikely in $N(=4)$ for the Standard Model Higgs to relax to the electroweak vacuum in the early universe.
\end{abstract}

\maketitle

\section{Introduction}

An important discovery in fundamental physics was that of the Higgs particle \cite{ATLAS,CMS}, which completes the Standard Model. Without the Higgs particle, the Standard Model would violate unitarity at energies well above the electroweak scale. The presence of the Higgs provides better UV behavior, allowing the Standard Model to make sense up to much higher energies. Interestingly, at temperatures well above the electroweak scale $\mathcal{O}(100)$\,GeV the particles of the Standard Model are massless. Furthermore, there is an associated set of symmetries which become manifest. In particular, we know that interacting massless spin 1 particles, must be coupled to exactly conserved charges, which by the Noether theorem are associated with symmetries. At lower temperatures, these particles become massive due to interesting dynamics of the Higgs. This affects the symmetry structure in dramatic ways and will be examined rigorously in this paper. 

Our primary interest is the Higgs mechanism in the Standard Model of particle physics. Here the full gauge group $SU(3)_c\times SU(2)_L\times U(1)_Y$ is linearly realized at high temperatures. Then, due to the Higgs acquiring a non-zero vacuum expectation value (VEV), only the reduced gauge group $SU(3)_c\times U(1)_{em}$ is linearly realized at low temperatures. The physical consequences can be stated more directly as follows: the 3 massless spin 1 particles $W_{1,2,3}$ associated with the group $SU(2)_L$ at high temperatures become 3 massive spin 1 particles $W^{\pm}, Z$ at low temperatures, while fermions acquire a mass too. Beyond the Standard Model, there can be Higgs mechanisms associated with other gauge groups. One interesting possibility is associated with grand unification, such as $SU(5)$ \cite{Georgi:1974sy} or $SO(10)$ \cite{Fritzsch:1974nn}, etc, whose larger set of massless spin 1 particles become massive at temperatures well below the grand unification scale. Another possibility is associated with various hidden sector gauge groups, which are suggested by ideas in fundamental physics, including string theory and models of dark matter, etc, which may be associated with other Higgs sectors. Furthermore, within condensed matter physics, the phenomenon of superconductivity is associated with the photon of $U(1)_{em}$ acquiring a mass at very low temperatures. This superconductivity example carries some of the physics relevant to elementary particle physics, despite the latter being expanded around the Lorentz invariant vacuum. In fact we will study the (Lorentz invariant) abelian $U(1)$ example often in this paper to illustrate the underlying physics in the simplest possible setting.

Now it is usually stated that the Higgs mechanism is responsible for the ``spontaneous breaking" of symmetries at low temperatures. The definition of spontaneous symmetry breaking (SSB) is that the theory is invariant under some symmetry $\hat{S}$, but the ground state (or the low temperature phase) of the theory is not, i.e., $\hat{S}|vac\rangle=|vac'\rangle$, where $|vac'\rangle\neq|vac\rangle$. While this is certainly true for some systems, such as ferromagnets, which spontaneously break rotational symmetry, it is rather unclear that this is correct for the Higgs mechanism where we need to deal with ``gauge symmetries". For one thing, consider the so-called ``small-gauge" transformations associated with functions $\alpha(x)$ that vanish at spatial infinity. These are not in fact actual symmetries but are merely redundancies in the mathematical description and can be entirely removed by gauge fixing. Hence they can never be spontaneously broken since all states in a theory are trivially invariant under a small-gauge transformation $|\psi'\rangle=|\psi\rangle$ (modulo an unimportant phase), including the vacuum. This is summarized by the so-called Elitzur's theorem \cite{Elitzur:1975im}. We will see this again later in this paper.

So the interesting and non-trivial issue to focus on is the fate of actual symmetries in a gauge theory. These are always associated with functions $\alpha(x)$ that are non-vanishing at infinity. The central example is when $\alpha(x)$ is a constant. This is a ``global" transformation, and is a sub-group of the full gauge group. These are sometimes referred to as ``global-gauge" transformations. (In the limit in which we send the gauge coupling to zero, these are ordinary global transformations.) If the Higgs has not acquired a VEV and so there are only massless spin 1 particles, these global transformations do in fact take generic states to different states $|\psi'\rangle\neq|\psi\rangle$ (at least for weakly coupled theories). However the vacuum itself is invariant under this symmetry transformation. (The vacuum may not be invariant under other symmetry transformations, such as the so-called ``large-gauge" transformations, in which $\alpha(x)$ is non-vanishing at infinity but also non-constant; we will return to these later in the paper.) 

The important question then is: {\em are the global symmetries spontaneously broken in the Higgs phase?} The answer does not seem obvious (and is not addressed by Elitzur's theorem \cite{Elitzur:1975im}). These are real symmetries at high temperatures in the regular phase (associated with conserved charges), but appear to be hidden at low temperatures. But does it mean these symmetries were spontaneously broken and there are many distinct vacua, or could it be that these symmetries were removed altogether from the theory and there is a unique vacuum? The answer to this question does not arise trivially from claiming that  gauge symmetry is only a redundancy, as this is ordinarily only true of the small-gauge transformations. For the fate of global symmetries, it requires a clear computation of the set of micro-states of the theory; we shall do this systematically in this paper. In particular, we will (i) explicitly count the vacuum states of the theory by finding the set of states $|vac\rangle$ that are orthogonal to one another. We will also (ii) count all the states in the Hilbert space by identifying the correct measure on the quantum path integral. 

There has been a long discussion in the literature on the topic of SSB in the context of gauge theories. This includes Refs.~\cite{Englert:1964et,Higgs:1964pj,Guralnik:1964eu,Bernstein:1974rd,Strocchi:1977za,Stoll:1995yg,Maas:2012ct,Kibble:2014gug,Strocchi:2015uaa,Maas:2017wzi} and references therein. While there is broad agreement on the physical character of the Higgs mechanism and the resulting predictions for experiment, there does remain some confusion on the important issue of whether SSB in gauge theories is physical or just nomenclature. In particular, there does not appear to be a direct computation of the overlap wave-function between ground states, and the associated complete characterization of the interpolation between the global and gauge cases. This overlap is the precise quantity required to sharply address the question of SSB, and will be performed explicitly in this paper (among other items). 

To calculate this (i) we will construct the ground state wave-functionals in the theory in the field basis. To work with a solvable system; we take the Higgs to be heavy, and so it can be taken to be frozen at its VEV $v$, as it has suppressed fluctuations. For the abelian $U(1)$ Higgs model, this leaves a truncated theory with only a quadratic Lagrangian in the transverse and longitudinal modes of a single spin 1 particle (for simplicity we ignore fermions, though their inclusion is straightforward). The longitudinal mode can be examined in various gauges, such as in Coulomb gauge $\nabla\cdot{\bf A}=0$ in which we must track a would-be Goldstone boson $\theta(x)$, or in unitary gauge $\theta(x)=0$ in which we must track the longitudinal part of the vector potential $A_L(x)$ directly. We show explicitly that in any gauge the ground state wave-functional is trivially invariant under small-gauge transformations, in accord with Elitzur's theorem. More interestingly, we then consider global transformations. We show that if the gauge coupling $g$ is set to zero, we obtain a family of orthogonal ground state wave-functionals in the large volume limit and so there is SSB. Conversely, for finite gauge coupling, we obtain a unique ground state wave-functional in the large volume limit and so there is no SSB. We also show that for finite volume, there is a smooth transition between these limiting cases as we change the gauge coupling away from zero. In particular, so long as the spin 1 particle's Compton wavelength $\lamC=2\pi/m=2\pi/(g\,v)$ is much bigger than the size of the system, we have SSB, but when it is much smaller than the size of the system, we have no SSB, and when it is comparable to the size of the system, we have partial SSB. 

In order to explain these results, we consider the structure of the global symmetry operator $\hat{S}$. In the quantum theory it is associated with a conserved charge operator $\hat{Q}$ as $\hat{S}=\exp(i\,\hat{Q}\,\theta_0)$. If the gauge coupling is taken to zero, we note there is a good conserved charge. However, for finite gauge-coupling the situation is more complicated. Recall that in a gauge theory, the charge $Q$ is given by a closed surface integral of the electric field $Q=\oint d{\bf S}\cdot{\bf E}$. For massless spin 1 particles, the electric field falls off as $1/r^2$ from point sources, leading to a conserved charge and a global symmetry. In fact there is a family of other charges defined at spatial infinity associated with large-gauge symmetries. However, when the spin 1 particle acquires a mass by the Higgs mechanism, the electric field falls off exponentially from point sources and so these charges are zero when the surface is taken to infinity and we show they are time dependent for finite surfaces. So there are no good conserved charges and no global (or large-gauge) symmetries. Hence the Higgs mechanism is in fact the removal of conserved charges and symmetries from the theory, rather than the spontaneous breakdown. By identifying canonically normalized fields, we also show this clearly at the level of the Lagrangian.

An interesting issue then is how to understand the existence and structure of defects, which are often thought to arise from SSB. For example, in the abelian $U(1)$ Higgs model, it is known that cosmic string solutions exist, and are usually ascribed to a phase-field that winds around the central axis of the string continuously varying from one vacuum to another. While this is the correct picture in the limit in which the gauge coupling is taken to zero, we explain that the basic properties of these cosmic strings at finite gauge coupling are best understood in terms of fields relaxing to a unique vacuum.

Another interesting issue is (ii) to count the full set of states in the theory beyond the vacuum states. In particular, since there is a unique vacuum and since in unitary gauge the would-be Goldstone modes $\theta(x)$ are removed from the theory, one might wonder what should be the measure on the quantum path integral. Should it be a volume measure $\mathcal{D}\rho\,\rho^{N-1}$, for the $N$ components of the Higgs multiplet, as it is in the global case, or should it be a line measure $\mathcal{D}\rho$ since the angular modes can be removed? By demanding that the theory is renormalizable, we prove that it is in fact the volume measure. This shows that in a definite sense the size of the Hilbert space in the global and gauge cases is in fact the same; it is simply that a symmetry gets removed in the gauge case.

An important application is to cosmology. We review the well known idea that the Higgs potential of the Standard Model turns over and becomes negative at very high energies, perhaps around $E\sim 10^{11}$\,GeV, or so, depending on parameters. We then utilize the volume measure to estimate the probability that the Higgs began on the favorable side of the potential in the early universe in order to roll down to our electroweak vacuum. We find that it is exponentially unlikely. However, this should be understood more carefully in the context of inflation, reheating, etc, which we comment on briefly.

Our paper is organized as follows:
In Section \ref{AbelianHiggsTheory} we construct the Lagrangian for the vector and scalar modes in a simple Higgs model.
In Section \ref{GroundStateWavefunctionals} we explicitly compute the ground state wave-functional(s).
In Section \ref{NumberVacua} we determine the number of vacua, comparing global to gauge cases.
In Section \ref{RealGauge} we examine whether the global transformations are real or redundant.
In Section \ref{Spectrum} we discuss the spectrum of the theory.
In Section \ref{Charges} we discuss the behavior of charges in the two phases.
In Section \ref{NonAbelian} we extend our results to non-abelian theories. 
In Section \ref{LargeGauge} we discuss large-gauge transformations.
In Section \ref{TopologocalDefects} we discuss properties of defects.
In Section \ref{PathIntegralMeasure} we derive the measure on the quantum path integral.
In Section \ref{ApplicationtoCosmology} we apply this to the Standard Model Higgs in the very early universe.
Finally, in Section \ref{Discussion} we present an outlook.

\section{Simple Higgs Theory}\label{AbelianHiggsTheory}

Our interest is in general Higgs theories, which may involve some non-abelian gauge groups. In the Standard Model this involves the non-abelian gauge group $SU(2)_L\times U(1)_Y$ at high temperatures (with the residual gauge group $U(1)_{em}$ at low temperatures). A complete analysis of this would involve the full treatment of the 3 massless bosons associated with $SU(2)_L$, which include self-interactions. However, this involves additional complications that are not essential for the main issues of state counting that we wish to analyze in this paper. Instead we will focus on a simple version of the Higgs mechanism, which involves the abelian gauge group $U(1)$ at high temperatures (with no residual group at low temperatures); see Section \ref{NonAbelian} for the non-abelian case. This in fact is an accurate description of superconductivity, and will highlight the important features of the Higgs in general settings.

Consider then a collection of identical spin 1 particles, ``photons", that are massless at high temperatures. To describe their features in a local way, we organize them into a vector field $A_\mu$, with associated field strength $F_{\mu\nu}=\partial_\mu A_\nu-\partial_\nu A_\mu$. Since massless spin 1 particles have only 2 (transverse) polarizations, we must ensure that 2 of the 4 components of $A_\mu$ are unphysical. This is readily achieved by making $A_0$ non-dynamical and by making another component redundant, i.e., by introducing a ``gauge symmetry". We minimally couple this field to a complex scalar $\cphi$ in the standard way with the following Lagrangian density (units $\hbar=c=1$, signature + - - -)
\be
\mathcal{L} = -\dfrac{1}{4}F_{\mu\,\nu}F^{\mu\,\nu} + |D_{\mu}\cphi|^2 - V(\abs{\cphi}),
\label{eq:Lagr_gaugedU1_regular}
\ee
where the covariant derivative is 
\be
D_\mu=\partial_\mu-i\,g\,A_\mu,
\ee
and $g$ is the gauge coupling. The complex scalar $\cphi$ implements the Higgs mechanism. It is assumed to have a potential of the form
\be
V(|\cphi|) = -\mu^2|\cphi|^2+\lambda|\cphi|^4,
\ee
where we have truncated all operators at the dimension 4 level (we will return to this issue in Section \ref{ApplicationtoCosmology} when we examine the potential more carefully). Note that the form of the Lagrangian density is unchanged under the familiar abelian gauge transformations with parameter $\alpha$ 
\bea
A_\mu & \to & A_\mu+\partial_\mu\alpha, \label{GaugeTransformationA} \\
\cphi & \to & \cphi\,e^{i\,g\,\alpha} \label{GaugeTransformationPhi} .
\eea
If $\alpha$ is a constant, this is a global transformation and will be studied very carefully in this paper; if $\alpha$ is non-constant and has support at infinity, this is a large-gauge transformation and will be studied briefly in this paper; if $\alpha$ is non-constant and vanishes at infinity, this is a small-gauge transformation and is merely a redundancy. 

In the regular phase $\mu^2<0$ leading to a minimum at $\cphi=0$. In the Higgs phase $\mu^2>0$ leading to a minimum at $\cphi\neq0$. To expand around the non-zero minimum in the Higgs phase, it is useful to decompose the field in terms of polar variables, with radial field $\rho$ and phase $\theta$ as
\be
\cphi({\bf x})={1\over\sqrt{2}}\,\rho({\bf x})\,e^{i\,\theta({\bf x})}.
\ee
Re-writing the Lagrangian density in terms of these polar variables gives
\be
\mathcal{L} = -\dfrac{1}{4}F_{\mu\,\nu}F^{\mu\,\nu} + {1\over2}(\partial_{\mu}\rho)^2 +  {1\over2}\rho^2(g\,A_{\mu} - \partial_{\mu}\theta)^2 - V(\rho),
\label{eq:Lagr_gaugedU1_euler}
\ee
and the corresponding potential for the radial field $\rho$ is
\be
V(\rho) = -\dfrac{1}{2}\mu^2\rho^2 + \dfrac{1}{4}\lambda\,\rho^4.
\label{HiggsPotential}\ee

Focussing on the Higgs phase with $\mu^2>0$, the minimum energy configuration occurs at the VEV $\rho_0=v=\mu/\sqrt{\lambda}$. We expand the radial Higgs field around this as
\be
\rho(x) = v + h(x),
\ee
and we refer to $h$ as the Higgs field, with associated quanta ``Higgs particles". This expansion leads to a Lagrangian density that decomposes into several pieces as follows
\be
\mathcal{L} = \mathcal{L}_{A_\mu,\theta} + \mathcal{L}_{h} + \mathcal{L}_{\text{int}},
\ee
where
\bea
\mathcal{L}_{A_\mu,\theta}  &=& -\dfrac{1}{4}F_{\mu\,\nu}F^{\mu\,\nu} + \dfrac{v^2}{2}(g\,A_{\mu} - \partial_{\mu}\theta)^2, \\
\mathcal{L}_{h} & = &{1\over2}(\partial_{\mu}h)^2 - \dfrac{1}{2}\,\lambda\,v^2\,h^2, \\
\mathcal{L}_{\text{int}} & = & \dfrac{1}{2}(2\,v\,h+h^2)(g\,A_{\mu} - \partial_{\mu}\theta)^2- \lambda\,v\,h^3-{1\over4}\lambda\,h^4.\,\,\,\,
\label{eq:Lagr_gaugedU1_euler_higgs}
\eea
Here $\mathcal{L}_{A_\mu,\theta}$ is the quadratic kinetic term for the photon and the would-be Goldstone $\theta$, $\mathcal{L}_h$ is the quadratic kinetic term for the Higgs, and $\mathcal{L}_{\text{int}}$ is the cubic and quartic interaction terms.

Note that the spin 1 mass and the Higgs mass are
\bea
m_A & = & g\,v,\\
m_h & = & \sqrt{2\,\lambda}\,v.
\eea
On the one hand, these are both set by the VEV $v$. On the other hand, they are parametrically different from one another, since the former is proportional to the gauge coupling $g$ and the latter is proportional to the (square-root of) self-coupling $\lambda$. Hence we can consider a situation in which the Higgs is somewhat heavier than the spin 1 particle (in the Standard Model the measured mass of the Higgs $m_h\approx 125$\,GeV \cite{Aad:2015zhl} is only a little heavier than the $W$ boson $m_W\approx 80$\,GeV \cite{Beringer:1900zz}). In this situation it suffices to consider the Higgs effectively frozen at its VEV, ignoring the back-reaction from its fluctuations. This means that the interaction terms can be ignored, and we can focus on the quadratic kinetic term $\mathcal{L}_{A_\mu,\theta}$ for the purpose of understanding the symmetry structure of the theory. This is precisely what makes the abelian $U(1)$ Higgs model so simple. In non-abelian cases, we would further need to consider interactions among the spin 1 particles. However we believe all our qualitative conclusions carry over to these more complicated situations.

Now an important feature of any unitary theory is that the time component of the vector potential $A_0$ is non-dynamical. By varying the above action with respect to $A_0$ we obtain the local version of Gauss' law
\be
\nabla\cdot{\bf E} = -\nabla^2A^0-\nabla\cdot\dot{\bf A}=-g\,v^2(g\,A^0-\dot\theta).
\label{GaussLaw}\ee
We can use this to solve for $A^0$. Since the reduced theory $\mathcal{L}_{A_\mu,\theta}$ is quadratic, we can diagonalize the theory in $k$-space. We define the spatial Fourier transform of a function $f$ as
\be
f_{\bf k}(t) \equiv \int \!d^3x\,e^{i\,{\bf k}\cdot{\bf x}} f({\bf x},t).
\ee
Then the solution for $A^0_{\bf k}$ is
\be
A^{0}_{\bf k} = \dfrac{i\,{\bf k}\cdot\dot{\bf A}_{\bf k} + g\,v^2\,\dot{\theta}_{\bf k}}{\omega_k^2},
\label{Azero}\ee
where the dispersion relation is
\be
\omega_k=\sqrt{k^2+g^2\,v^2},
\label{Dispersion}\ee
with $k\equiv|{\bf k}|$.

Having eliminated $A^0$, there are 4 dynamical fields remaining in $\mathcal{L}_{A_\mu,\theta}$, namely $A_1,A_2,A_3,\theta$. However there are only 3 physical degrees of freedom as there is a gauge redundancy between them. For these 3 physical degrees of freedom, it is useful to isolate the 2 transverse vector modes, and the 1 longitudinal scalar mode. To do so we define a transverse vector ${\bf A}^T_{\bf k}$ and a longitudinal scalar $A^L_{\bf k}$ as
\bea
{\bf A}^T_{\bf k} & = & {\bf A}_{\bf k}-\hat{\bf k}\,(\hat{\bf k}\cdot{\bf A}_{\bf k}), \\
A^L_{\bf k} & = & i\,\hat{\bf k}\cdot{\bf A}_{\bf k},
\eea
where $\hat{\bf k}\equiv{\bf k}/k$ is a unit vector in the ${\bf k}$-direction.

By using the solution for $A^0$ in $k$-space, we find that the Lagrangian for the transverse (T) and longitudinal (L) modes decouples and can be written as
\be
L_{A_\mu,\theta} = \int\!\dfrac{d^3k}{(2\pi)^3}\left(\mathcal{L}_{T,\bf k} + \mathcal{L}_{L,\bf k}\right),
\ee
where
\bea
\mathcal{L}_{T,\bf k} & = & {1\over2}|\dot{\bf A}^T_{\bf k}|^2-{1\over2}\omega_k^2|{\bf A}^T_{\bf k}|^2, \label{LagTrans}\\
\mathcal{L}_{L,\bf k} & = & \dfrac{v^2}{2\,\omega^2_k}|g\,\dot{A}^L_{\bf k} + k\,\dot{\theta}_{\bf k}|^2 - \dfrac{v^2}{2}|g\,A^L_{\bf k} + k\,\theta_{\bf k}|^2. \label{LagLong}
\eea

Note that this Lagrangian is invariant under the (local) gauge transformations Eqs.~(\ref{GaugeTransformationA},\,\ref{GaugeTransformationPhi}), since the fields transform as
\bea
{\bf A}^T_{\bf k} & \to & {\bf A}^T_{\bf k}, \label{ATgauge}\\
A^L_{\bf k} & \to & A^L_{\bf k}-k\,\alpha_{\bf k}, \label{ALgauge}\\
\theta_{\bf k} & \to & \theta_{\bf k}+g\,\alpha_{\bf k}, \label{Thetagauge}
\eea
which trivially leaves $\mathcal{L}_T$ unchanged and also leaves $\mathcal{L}_L$ unchanged as $g\,A^L_{\bf k} + k\,\theta_{\bf k}\to g\,A^L_{\bf k} + k\,\theta_{\bf k}$. It will be useful to analyze this theory in various gauges, especially unitary gauge $\theta_{\bf k}=0$ and Coulomb gauge $A^L_{\bf k}=0$ in which the form of Eq.~(\ref{LagLong}) appears quite different.

\section{Ground State Wave-functional(s)}\label{GroundStateWavefunctionals}

Among other things, we will be interested in the ground state of the theory. Since the above theory is quadratic, it is really a set of harmonic oscillators, whose ground state wave-function are Gaussians. To make the structure even clearer, let us pass to the Hamiltonian. It again decouples into a transverse (T) and longitudinal (L) piece as
\be
H_{A_\mu,\theta}= H_T+H_L,
\ee
where
\bea
H_{T} & = & \int\! \dfrac{d^3k}{(2\pi)^3}\left[\dfrac{1}{2}|{\vec\Pi}^T_{\bf k}|^2 + \dfrac{1}{2}\omega^2_k|{\bf A}^T_{\bf k}|^2\right], \label{eq:H_spin1}\\
H_{L} & = & \int\! \dfrac{d^3k}{(2\pi)^3}\left[\dfrac{\omega^2_k}{2\,v^2}|{\Pi}^L_{\bf k}|^2 + \dfrac{v^2}{2}|g\,A^L_{\bf k} + k\,\theta_{\bf k}|^2\right], \label{eq:H_spin0}
\eea
with momentum conjugates
\bea
{\vec\Pi}^T_{\bf k} & = & \dot{\bf A}^{T*}_{\bf k}, \\
{\Pi}^L_{\bf k} & = & {v^2\over\omega^2_k}(g\,\dot{A}^L_{\bf k} + k\,\dot{\theta}_{\bf k})^*.
\eea
Since the Hamiltonian is additive in the transverse and longitudinal modes, the ground state wave-functional $\Psi$ factorizes as
\be
\Psi(A_\mu,\theta) = \Psi_T({\bf A}^T_{\bf k})\,\Psi_L(A^L_{\bf k},\theta_{\bf k}).
\ee

Now recall that for a single harmonic oscillator with standard Hamiltonian $H=p^2/(2m)+m\,\omega^2\,x^2/2$, the ground state wave-function in the $x$-basis is $\psi(x)\propto\exp(-m\,\omega\,x^2)$. Similarly we can determine the ground state wave-functionals for the above Hamiltonians in the field basis to be
\bea
\Psi_T({\bf A}^T_{\bf k}) &\propto& \exp[-{1\over2} \int\!\dfrac{d^3k}{(2\pi)^3}\,\omega_k\,|{\bf A}^{T}_{\bf k}|^2],\label{eq:ground_stateT}\\
\Psi_L(A^L_{\bf k},\theta_{\bf k}) &\propto& \exp[-{1\over2} \int\!\dfrac{d^3k}{(2\pi)^3}\dfrac{v^2}{\omega_{k}}|g\,A^{L}_{\bf k} + k\,\theta_{\bf k}|^2].
\label{eq:ground_stateL}
\eea
Since it is not important, we will not explicitly report on the normalization factors here, but they can be readily determined, and our final results will properly include this.

Note that under (local) small-gauge transformations Eqs.~(\ref{ATgauge},\,\ref{ALgauge},\,\ref{Thetagauge}), the ground state wave-functional is left exactly invariant (in fact it does not even pick up a phase). This is trivially expected, as {\em all} states in a theory are invariant under (local) small-gauge transformations (modulo an unimportant phase), as these are mere redundancies in the description. This is in accord with Elitzur's theorem \cite{Elitzur:1975im}. 

The non-trivial issue is the behavior of the ground state wave-functionals under a global transformation. In this case we should take the function $\alpha({\bf x})$ to be a constant. However, understanding its behavior is somewhat confusing in $k$-space. This is because in $k$-space, we would formally have $\alpha_{\bf k}\propto \delta^3({\bf k})$, meaning we have to be extremely careful about the infrared behavior of modes. 

So in order to regulate the infrared in a clear fashion, it is useful to define the theory in a box of volume $V$ and study the ground state wave-functional in position space. To illustrate the idea, it is convenient (though not necessary) to pick a gauge. In unitary gauge $\theta=0$ and the wave-functional $\Psi_L$ appears to obviously only have a unique vacuum centered around $\langle A^L\rangle=0$. On the other hand, the situation is less clear in Coulomb gauge $A^L=0$, since it may appear that there is a family of vacua labelled by different choices of $\langle\theta\rangle=\theta_0$. 

So to address this confusion, let us proceed to operate in Coulomb gauge, and compute $\Psi_L$ in position space. Starting with Eq.~(\ref{eq:ground_stateL}) we Fourier transform and the wave-function is readily found to be
\be
\Psi_L(\theta)\propto\exp\left[-{1\over2}\int_V\! d^3x\int_V d^3x'\, \theta({\bf x})\, \K_\uv({\bf x}-{\bf x}')\, \theta({\bf x}')\right],
\label{eq:ground_stateL_position}\ee
where $\K_\uv$ is the following kernel
\be
\K_\uv(r) = v^2\!\int\!{d^3k\over(2\pi)^3}{k^2\over\omega_k}\,e^{-i\,{\bf k}\cdot\,{\bf r}}\,e^{-k\,\uv}.
\label{Kernel}\ee
Here $\uv$ is a UV regulator that we take towards zero at the end of the computation in order to have well defined quantitites. If the field theory is defined on the lattice, then $\uv$ roughly sets the lattice spacing. 

For a compact angular variable, we should in fact sum this wave-functional to obtain the total wave-functional $\tilde{\Psi}_L$ as 
\be
\tilde\Psi_{L}(\theta) = \sum_n\Psi_L(\theta+2\pi\,n),
\ee
for integers $n$ to ensure periodicity under $\theta\to\theta+2\,\pi\,n$.

\section{Number of Vacua}\label{NumberVacua}

We wish to determine if there are in fact many ground state wave-functionals. Under a global transformation, $\theta({\bf x})$ gets shifted by a constant as
\be
\theta({\bf x}) \to \theta({\bf x})+\theta_0.
\ee
Such a transformation definitely leaves the Lagrangian and indeed the Hamiltonian unchanged. Hence this transforms the above ground state wave-functional to another ground state wave-functional
\be
\Psi(A_\mu,\theta)\to\Psi'(A_\mu,\theta)=\Psi(A_\mu,\theta+\theta_0).
\ee

However we must be careful to check if this is really a new state or if it is just a re-writing of the old state. To check, we need to compute the overlap between the original vacuum $|\vac\rangle$ and the transformed vacuum $|\vac'\rangle$ 
\be
\langle \vac|\vac'\rangle = {\int\!\mathcal{D}\theta\,\tilde\Psi_L(\theta)\,\tilde\Psi_L(\theta+\theta_0)\over \int\!\mathcal{D}\theta\,\tilde\Psi_L(\theta)\,\tilde\Psi_L(\theta)},
\ee
where we have divided out by the appropriate normalization factor and we have suppressed the integral over the transverse modes as this trivially cancels out. Since the wave-functional is a Gaussian Eq.~(\ref{eq:ground_stateL_position}), this overlap integral is straightforward to carry out and leads to a Gaussian in $\theta_0$, namely
\be
\langle \vac|\vac'\rangle \propto \sum_n\exp\left[-{1\over4}(\theta_0+2\,\pi\,n)^2\,\,V\!\!\int_V d^3 r\, \K_\uv(r)\right].
\label{overlap}\ee
We have simplified the argument of the exponential by switching from ${\bf x}$ and ${\bf x}'$ to center of mass co-ordinates, integrated over the the center of mass co-ordinate $({\bf x}+{\bf x}')/2$ giving the volume factor $V$, leaving an integral over the relative co-ordinate ${\bf r}={\bf x}-{\bf x}'$. 

Note that if $\int_V d^3r\,\K_\uv(r)$ was a constant, independent of $V$, then Eq.~(\ref{overlap}) would say that the overlap between the original state and the transformed state is exponentially small in the volume of space. Hence we recover the well known idea that SSB in a quantum theory is a {\em large volume effect}, since $\langle \vac|\vac'\rangle\to0$ in the large volume limit. In fact if we were studying a double well potential, and we were comparing the overlap between states built on each of the wells, we would indeed find this was the case.

However we now need to study the integral $\int_V d^3r\,\K_\uv(r)$ carefully in the case at hand. This integral can be determined in the following way. The factor of $k^2$ in Eq.~(\ref{Kernel}) can be pulled out by acting with the Laplacian on a related kernel $\Kr$ as follows
\be
\K_\uv(r)=-\nabla^2\Kr_\uv(r),
\label{Kernelr}\ee
where 
\be
\Kr_\uv(r) = v^2\!\int\!{d^3k\over(2\pi)^3}{1\over\omega_k}\,e^{-i\,{\bf k}\cdot\,{\bf r}}\,e^{-k\,\uv}.
\ee
Using Eq.~(\ref{Dispersion}) for the dispersion relation $\omega_k$ allows this Fourier transform to be computed explicitly. For $r\gg\uv$, we find it is
\be
\Kr_\uv(r) = {g\,v^3\,K_1(g\,v\,r)\over2\,\pi^2\,r},\,\,\,\,\, r\gg\uv,
\label{Juv}\ee
where $K_1$ is the modified Bessel function of the second kind of order 1. Inserting Eq.~(\ref{Kernelr}) into Eq.~(\ref{overlap}) and using the divergence theorem allows us to express the argument of the exponent as the following boundary term
\be
\langle \vac|\vac'\rangle \propto \sum_n \exp\left[{1\over4}(\theta_0+2\,\pi\,n)^2\,\,V\!\!\oint d{\bf S}\cdot\nabla\Kr_\uv\right],
\label{overlapboundary}\ee
where $d{\bf S}$ is an infinitesimal area vector on the boundary of the spatial region. Note that since this is a boundary term, so long as the size of the boundary is much greater than our UV cutoff $\uv$, we can simply use the $r\gg\uv$ result in Eq.~(\ref{Juv}).

\begin{figure}[t]
\centering
\includegraphics[width=\columnwidth]{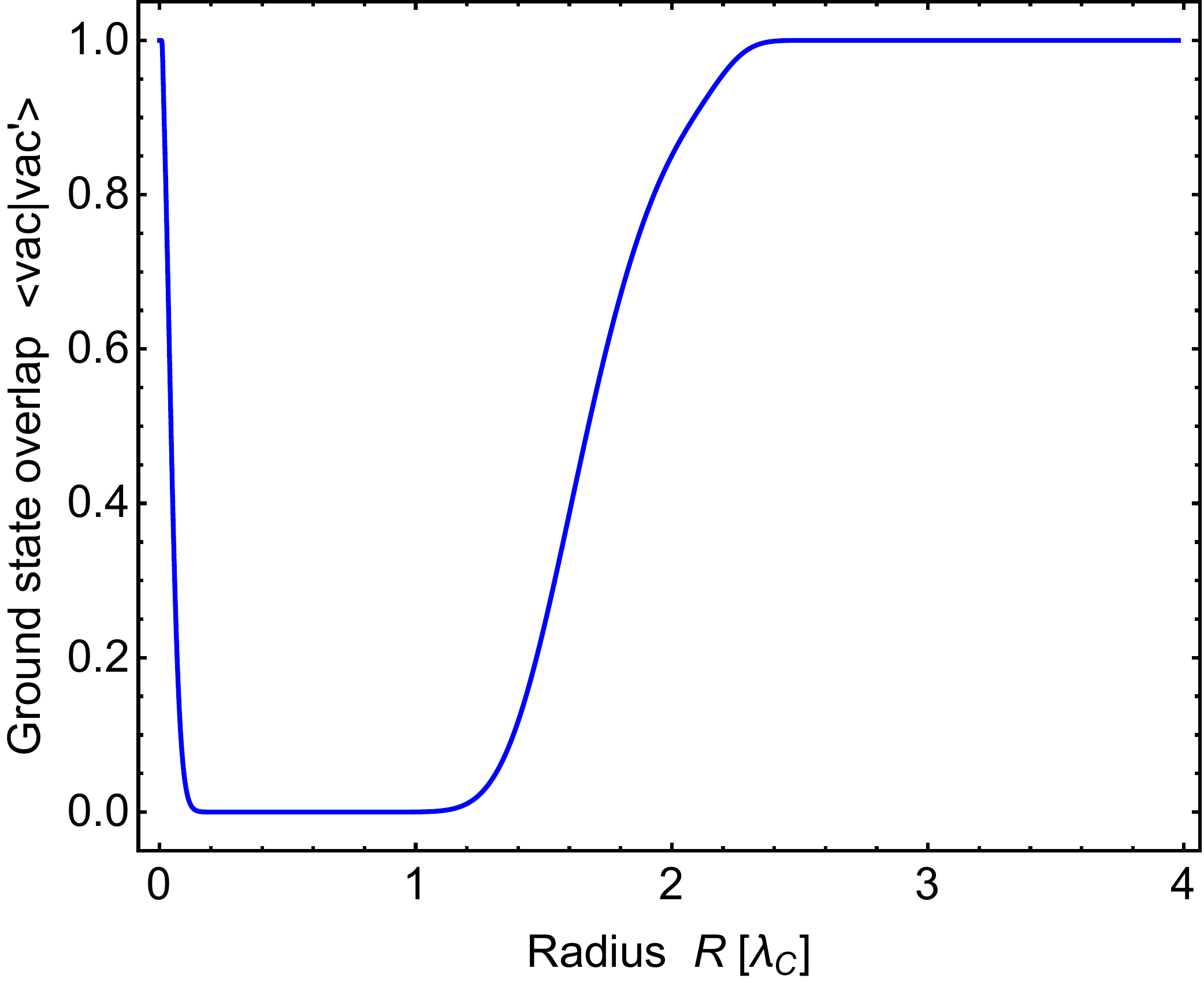}
\caption{The value of the ground state overlap $\langle\vac|\vac'\rangle$ as a function of radius $R$ (in units of $\lamC=2\pi/m_A=2\pi/(g\,v)$) according to the result in Eq.~(\ref{overlapexact}). We have set $g=0.3$ and $\theta_0=\pi/4$ here.}
\label{FigureOverlap}
\end{figure}

We can evaluate this explicitly by taking the spatial region to be a sphere of radius $R$. Carrying out the integral, taking $\uv\to 0$, computing the derivative of $J$, and writing out the normalization factor explicitly, leads to an important result
\be
\langle \vac|\vac'\rangle = {\sum_n\exp\left[-{2\over3}\,(\theta_0+2\,\pi\,n)^2\,g^2\,v^4\,R^4\,K_2(g\,v\,R)\right]\over \sum_n\exp\left[-{2\over3}\,(2\,\pi\,n)^2\,g^2\,v^4\,R^4\,K_2(g\,v\,R)\right]},
\label{overlapexact}\ee
where $K_2$ is the modified Bessel function of the second kind of order 2. This result can be re-written in terms of the elliptic theta function of order 3, but we suppress those details here. The behavior is shown in Fig.~\ref{FigureOverlap}.

We can analyze this overlap in two important cases depending on the Compton wavelength of the massive spin 1 particle 
\be
\lamC={2\pi\over m_A}={2\pi\over g\,v}.
\ee
Case (a) is when $\lamC$ is much larger than $R$, and case (b) is when $\lamC$ is much smaller than $R$. In these two limits, the value of the modified Bessel function is very different. In case (a) $K_2(x)\to 2/x^2$ leading to the overlap being exponentially small
\be
\langle \vac|\vac'\rangle = e^{-4\,\theta_0^2\,v^2\,R^2/3},\,\,\,\,\,\lamC \gg R,
\ee
where we have used the fact that the $n=0$ term dominates when the argument here is large. This behavior is seen in the left region of Fig.~\ref{FigureOverlap}. While in case (b) $K_2(x)\to\sqrt{\pi/(2\,x)}\,\exp(-x)$ leading to the overlap rapidly approaching unity
\be
\langle \vac|\vac'\rangle = 1,\,\,\,\,\,\lamC \ll R,
\ee
(the correction from 1 is in fact doubly exponentially small at large $R$).
This behavior is seen in the right region of Fig.~\ref{FigureOverlap}. 

Case (a) is that of an ordinary global theory, obtained by sending the gauge coupling $g$ to zero. In this limit $\lamC\to\infty$, guaranteeing that any finite size sphere trivially satisfies $\lamC\gg R$. We see that the overlap goes to zero as an exponential in $R^2$. This is not as fast as an exponential fall off with volume $V\sim R^3$, but is still rather fast. Hence this recovers the well known idea that in an un-gauged field theory, with a global shift symmetry $\theta({\bf x})\to\theta({\bf x})+\theta_0$, there is an orthogonal set of vacua labelled by different values of $\langle\theta\rangle=\theta_0$ in the large volume limit; which is SSB.

On the other hand, case (b) is that of an ordinary gauge theory with finite gauge coupling $g$. For $W$ and $Z$ bosons in the Standard Model, the Compton wavelength is $\lamC\sim 10^{-17}$\,m. So as long as one is considering spatial regions of size much bigger than $10^{-17}$\,m, which is usually the case, then we are in this regime. Here the behavior is radically different. Instead of the overlap wave-functional approaching zero exponentially fast, it approaches one exponentially fast. This means that in the large volume regime, all these wave-functionals are in fact the same. This means there is only a single unique vacuum and no SSB.

We note that the overlap result in Eq.~(\ref{overlapexact}) provides a smooth interpolation between these limiting cases. There is at least one situation in which this full result may be important. Instead of heavy $W$ and $Z$ bosons, consider the photon itself. Now it is likely that the photon is strictly massless, but we do not know for sure. A claimed observational bound on the photon mass is $m_\gamma\lesssim10^{-26}$\,eV from galactic magnetic fields \cite{Goldhaber:2008xy} (though the validity of this bound can be debated \cite{Adelberger:2003qx}). Suppose the photon does indeed have a non-zero mass near this upper bound. This corresponds to a very large Compton wavelength of $\lamC\sim$\,kpc. So if we are interested in physics on scales much larger than $\sim$\,kpc, then there is only a unique vacuum (right region of Fig.~\ref{FigureOverlap}). However, if we focus on physics on scales $\sim$\,kpc, then there is effectively many degenerate vacua (middle-left region of Fig.~\ref{FigureOverlap}). This is similar to the case of having an ultra-light scalar, where depending on the regime of interest, either it is sitting at its true unique vacuum or it may be displaced away from its true vacuum and could be sitting at one of many other effectively degenerate vacua.

\section{Real or Redundant Transformation?}\label{RealGauge}

Above we showed that at finite gauge coupling, there is no SSB in the large volume regime. We can understand this further by re-examining the kernel $\K(r)$ that determines the ground state wave-functional. Since the physics at large volume is controlled by small wave-numbers, we can gain some understanding by Taylor expanding the factor $k^2/\omega_k$ around $k=0$ as
\be
{k^2\over\omega_k}={k^2\over m_A}-{k^4\over2\,m_A^3}+\ldots.
\ee
By interchanging the order of summation and integration, we can readily compute $\K(r)$ as a series expansion. The $n^{th}$ term will be proportional to the $2n^{th}$ derivative of the delta-function, i.e.,
\be
\K(r)=-{v^2\over m_A}\left[\nabla^2\delta^3({\bf r})+{\nabla^4\delta^3({\bf r})\over2\,m_A^2}+\ldots\right].
\ee
This is related to the above exact result for the kernel; since the kernel is a modified Bessel function, it is large near $r\to 0$, but is exponentially small as $r\to\infty$, which is similar to the structure of the delta-function. By inserting this into the wave-functional in Eq.~(\ref{eq:ground_stateL_position}), integrating by parts, and integrating over ${\bf x}'$ using the delta-function, we obtain
\be
\Psi_L(\theta)\propto\exp\left[-{v^2\over2\,m_A}\int d^3x \left[(\nabla\theta)^2-{(\nabla^2\theta)^2\over2\,m_A^2}+\ldots\right]\right].
\ee
We have implicitly taken the box size to infinity here in order to integrate by parts and throw away all boundary terms. 

In this limit we see that the wave-functional is really only a function of $\nabla\theta$ and not $\theta$. Therefore it is trivially invariant under the transformation $\theta({\bf x})\to\theta({\bf x})+\theta_0$. This indicates that such transformation are really just pure gauge transformations (redundancies) in the gauged Higgs phase, even though they are real (global) transformations in the regular phase and in the un-gauged Higgs phase.

In fact we can understand this further by returning to the classical Lagrangian density for the longitudinal modes Eq.~(\ref{LagLong}). Again let us operate in Coulomb gauge $A^L=0$. Then let us re-write the Lagrangian density back in position space. Since there are complicated powers of wavenumber $k$, we need to utilize some formal operations involving inverse powers of Laplacians, etc, as follows
\be
\mathcal{L}_L = {v^2\over2}\nabla\dot\theta{1\over-\nabla^2+g^2\,v^2}\nabla\dot\theta-{v^2\over2}(\nabla\theta)^2.
\label{thetaLag}\ee
This is indeed a peculiar looking Lagrangian for a scalar field, which we now analyze. 

Firstly, note that in the limit in which we send the gauge coupling to zero, we recover the standard Lagrangian of a massless scalar field
\be
\mathcal{L}_L \to {v^2\over2}\dot\theta^2-{v^2\over2}(\nabla\theta)^2,\,\,\,\,\,\mbox{as}\,\,\,\,\,g\to0.
\ee
since the spatial derivatives in the first term of Eq.~(\ref{thetaLag}) formally cancels out. In this Lagrangian with $g\to0$ there is obviously an ordinary global shift symmetry $\theta({\bf x})\to\theta({\bf x})+\theta_0$ that is real and physical. It is associated with the existence of a massless boson, by the Goldstone theorem.

On the other hand, for finite gauge coupling, the Lagrangian density in Eq.~(\ref{thetaLag}) is rather different. Its radical departure from that of an ordinary massless scalar arises from the coupling to the gauge field and then solving and re-inserting for $A^0$. If we focus on low momenta, or small gradients, compared to the spin 1 particle's mass, we can Taylor expand in $|\laplacian|\ll g^2\,v^2=m_A^2$ as follows
\be
\mathcal{L}_L = {v^2\over2\,m_A^2}\left[(\nabla\dot\theta)^2-{(\laplacian\dot\theta)^2\over m_A^2}+\ldots\right]-{v^2\over2}(\nabla\theta)^2.
\label{thetaLag2}\ee
We now see that the Lagrangian density is trivially invariant under $\theta(x)\to\theta(x)+\theta_0$ since every term is only a function of $\nabla\theta$. In fact the Lagrangian density is invariant under a much larger set of transformations $\theta(x)\to\theta(x)+f(t)$, where $f(t)$ is an arbitrary function of time. Such transformations are not physical, but are in fact pure redundancies. This is directly connected to the above ground state wave-functional being also only a function of $\nabla\theta$ at finite gauge coupling, and being mapped into itself under a shift in $\theta$.

\section{Spectrum}\label{Spectrum}

In fact we can understand the behavior of the longitudinal mode much more directly by simply passing to a canonically normalized field. We can do this in any gauge, so let us now go back to a general gauge, and define the gauge invariant variable
\be
\phic({\bf x},t)\equiv \int\!{d^3k\over(2\pi)^3}{g\,A^L_{\bf k}(t)+k\,\theta_{\bf k}(t)\over\omega_k}\,e^{-i\,{\bf k}\cdot{\bf x}}.
\label{canonical}\ee
This new variable $\phic$ is described by the manifestly Lorentz invariant (as it is gauge invariant) Lagrangian density
\be
\mathcal{L}_L = v^2\!\left[{1\over2}\dot\phic^2-{1\over2}(\nabla\phic)^2-{1\over2}g^2\,v^2\,\phic^2\right],
\ee
for any value of the gauge coupling $g$. It is now clear to see that it is only in the limit in which we take $g=0$ do we obtain a physical shift symmetry $\phic(x)\to\phic(x)+\phiz$. Furthermore from Eq.~(\ref{canonical}) we note that it is only in this limit that $\phic=\theta$. Otherwise the canonically normalized field $\phic$ is very different from $\theta$. For non-zero gauge coupling, the field $\phic$ has a mass $g\,v$. This is of course the mass of the longitudinal (and transverse) modes of the massive spin 1 particle. This shows explicitly that for non-zero gauge coupling there is only a unique vacuum at $\phic=0$. If the gauge coupling is very small, and we focus on scales $R$ much less than the Compton wavelength of the particle, there is effectively a set of degenerate vacua, as we discussed in Section \ref{NumberVacua}.

\begin{figure}[t]
\centering
\includegraphics[width=\columnwidth]{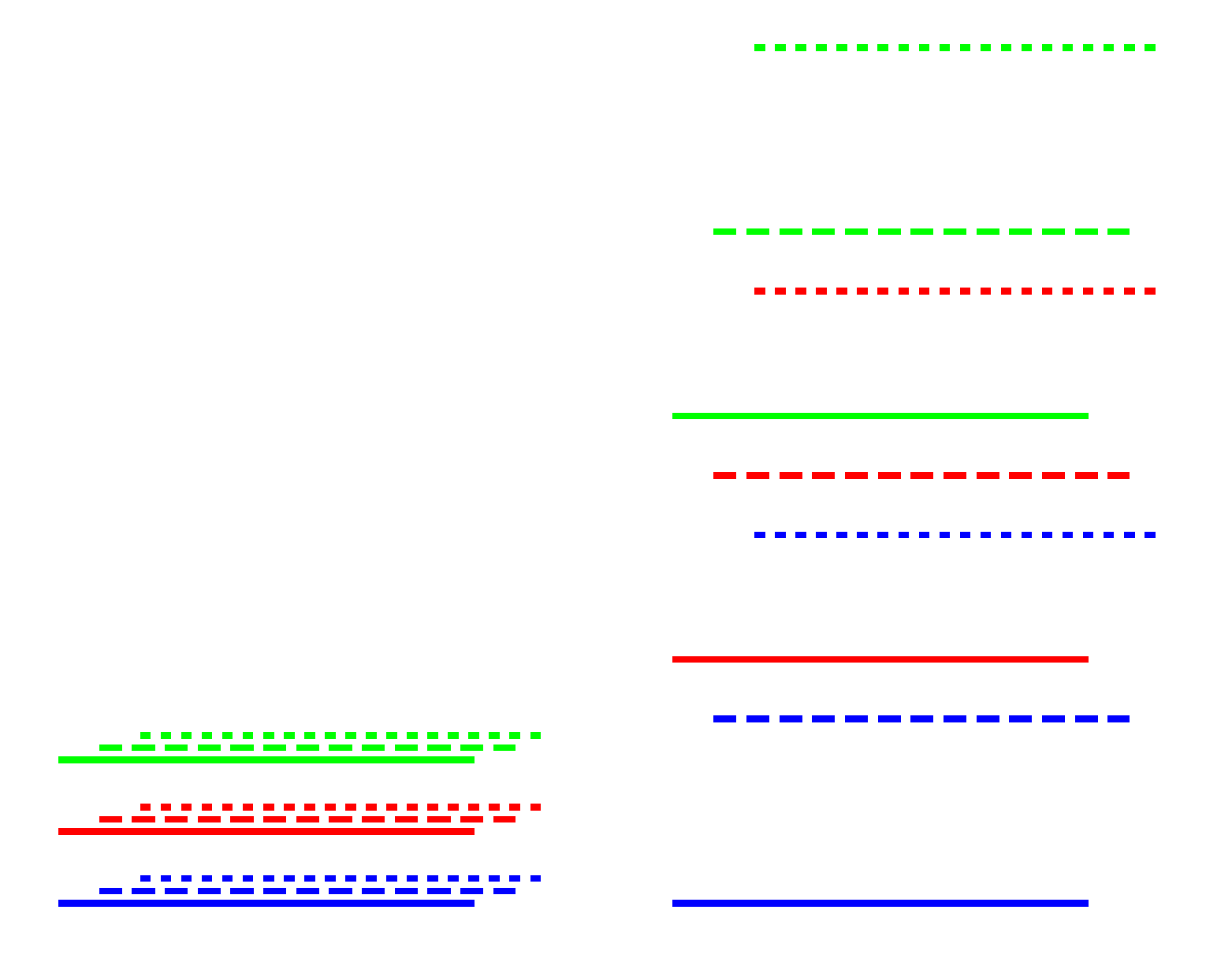}
\caption{A few of the low lying energy levels of the Hamiltonian. Left is the global case with $g=0$. Right is the gauge case with $g\neq0$. Note there is a one-to-one correspondence between the states in the two cases, as indicated by the colors and the line styles. In the global case there is degeneracy, while in the gauge case the degeneracy is lifted.}
\label{FigureSpectrumMassless}
\end{figure}

The key idea then is that in the gauged-Higgs theory the degeneracy of the global theory has been lifted, leaving a unique vacuum. We can see a direct connection between the set of states in the global case and the gauge case. We have illustrated a few of the low lying energy levels in Fig.~\ref{FigureSpectrumMassless}. The blue represents states of zero momentum, the red represents states with the lowest allowed non-zero momentum $k=2\pi/L$, and the green represents states of the second lowest allowed non-zero momentum $k=4\pi/L$. Solid blue curve represents the initial vacuum. Dashed represents adding a particle of zero momentum. Dotted represents adding another particle of zero momentum. There is a one-to-one mapping between the states of the global and gauge cases. So it is important to note that the reason there is only one unique vacuum in the gauge case is not because we have simply declared all the vacua to be equal and reduced the size of the Hilbert space by hand. In fact the size of the Hilbert space in the two cases is in some sense the same (we shall return to this in Section \ref{PathIntegralMeasure}). It is simply that the eigen-spectrum has been radically altered due to interesting dynamics.

\section{Charges}\label{Charges}

The above analysis shows that while in the regular phase there is a global symmetry, there is no such symmetry in the Higgs phase; it is removed from the theory, as opposed to being spontaneously broken. We can understand this further by analyzing the behavior of charges in the theory.

Recall that Gauss' law provides a boundary valued representation of the enclosed charge in a gauge theory, namely
\be
Q_{enc} = \int_V\! d^3x\,\nabla\cdot{\bf E} = \oint_S d{\bf S}\cdot{\bf E}.
\label{ChargeGauss}\ee
In the regular phase of a gauge theory, the electric field has a monopole piece from ordinary sources that satisfies $|{\bf E}|\propto 1/R^2$ at the boundary. This $1/R^2$ fall off is counter-balanced by the area of the boundary that scales as $S\propto R^2$, to give a total charge $Q_{enc}$ that is independent of $R$, finite, and conserved (unless charges cross the boundary).

However, in the Higgs phase of a gauge theory, the behavior of the electric field and charge is radically different. First we need to check if $Q_{enc}$ is a conserved quantity. To address this question, let us return to the form of Gauss' law in the case of the abelian Higgs model. Let us now include  charged matter to act as a source, whose contribution to the charge density we denote $\rho_m$. Suppose the source is a classical point charge $q$  located at position ${\bf x}_q$; its charge density is
\be
\rho_m({\bf x}) = q\,\delta^3({\bf x}-{\bf x}_q).
\ee
In unitary gauge the local form of the Gauss' law becomes
\be
\nabla\cdot{\bf E}=-\nabla^2A^0-\nabla\cdot\dot{\bf A} = -m_A^2 \,A^0+\rho_m,
\ee
where $m_A=g\,v$. For very slowly moving sources $v\ll c$ we can ignore the $\nabla\!\cdot\!\dot{\bf A}$ term as it is a relativistic correction. In this limit the solution for $A^0$ is simple to obtain
\be
A^0({\bf x}) \approx {q\,e^{-m_A|{\bf x}-{\bf x}_q|}\over4\,\pi\,|{\bf x}-{\bf x}_q|},
\ee
with corresponding electric field is given by
\be
{\bf E}({\bf x},0) = {q\left(1+m_A\,|{\bf x}-{\bf x}_q|\right)e^{-m_A\,|{\bf x}-{\bf x}_q|}\over 4\,\pi\,|{\bf x}-{\bf x}_q|^3}({\bf x}-{\bf x}_q).
\ee
The exponential fall off $|{\bf E}|\propto e^{-m_A\,R}/R$ for $R\gg 1/m_A$ is not counter-balanced by the area of the boundary $S\propto R^2$. So we now have a charge $Q_{enc}$ that depends sensitively on how close the source is to the boundary.

\begin{figure}[t]
\centering
\includegraphics[width=\columnwidth]{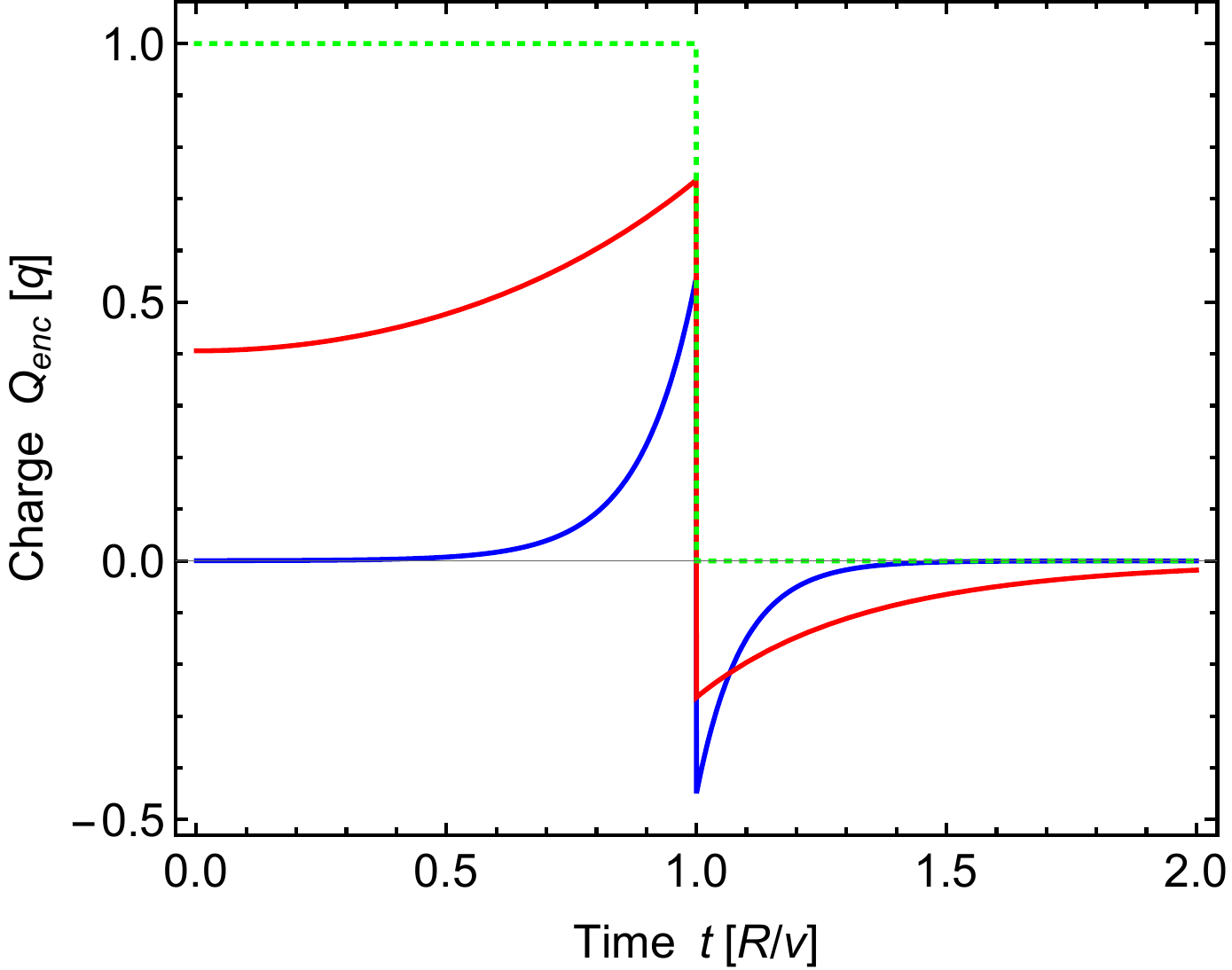}
\caption{The charge $Q_{enc}$ enclosed by a sphere of radius $R$ centered at the origin as a function of time $t$ ($Q_{enc}$ is in units of $q$ and $t$ is in units of $R/v$). The source is a classical point charge $q$ that slowly ($v\ll c$) moves radially outwards at constant speed with position $r_q=v\,t$. The three curves represent different choices for the spin 1 mass: solid blue is $m_A=10/R$, solid red is $m_A=2/R$, and dotted green is $m_A=0$.}
\label{FigureCharges}
\end{figure}

As a concrete example, we take the boundary to be a sphere of radius $R$ centered at the origin ${\bf x}=0$. By integrating the electric field over the sphere using Eq.~(\ref{ChargeGauss}) we obtain the enclosed charge to be
\be
Q_{enc} = \Bigg{\{}
\begin{array}{l}
{q\,e^{-m_A\,R}(1+m_A\,R)\sinh(m_A\,r_q)\over m_A\,r_q},\,\,\,\,\,\,\,\,\,\,\,\,\,\,\,\,\,\,\,\,\,\,\,\,\,\,\,\,\,\,\,r_q<R,\\
{q\,e^{-m_A\,r_q}(\sinh(m_A\,R)-m_A\,R\,\cosh(m_A\,R))\over m_A\,r_q},\,\,\,\,\,r_q>R,
\end{array}
\ee
(with $r_q\equiv|{\bf x}_q|$), where there is a change in behavior depending on whether the point source is inside or outside the sphere. Let us take the source to be moving very slowly away from the origin on a straight line with constant speed $v$ as
\be
r_q = v\,t.
\ee
So for times $0\leq t<R/v$ the source is inside the sphere, then for times $t>R/v$ the source is outside the sphere. The corresponding enclosed charge is plotted in Fig.~\ref{FigureCharges}. For non-zero gauge coupling $m_A=g\,v\neq 0$ we obtain the solid blue ($m_A=10/R$) and solid red ($m_A=2/R$) curves. In this case the enclosed charge is time dependent both inside and outside the sphere; so it is not conserved even when sources are not crossing the boundary. For zero gauge coupling $m_A=g\,v=0$ we obtain the dotted green curve. In this case we recover the usual result that the enclosed charge is exactly conserved, except when crossing the boundary.

On the other hand, for non-zero gauge coupling the charge is conserved in the limit that we send the size of the system $R\to\infty$ and consider only localized sources contained within this infinite sphere. However, although the charge becomes conserved in this limit, it only does so in a trivial sense; the charge itself approaches zero exponentially fast
\be
Q\to 0,\,\,\,\,\,R\to\infty.
\ee

Now recall that in the quantum theory, charges are the generators of symmetries
\be
\hat{S}=e^{i\,\hat{Q}\,\theta_0}.
\ee
In the regular phase, or in the zero gauge coupling case, the charge operator is non-zero and conserved, leading to a non-trivial symmetry operator $\hat{S}$ that acts on a generic state $|\psi\rangle$ and usually takes it to a new state
\be
|\psi'\rangle=\hat{S}|\psi\rangle\neq|\psi\rangle.
\ee
Although for some special states, namely states of definite charge, it maps the state to itself (modulo a possible phase).

However, in the Higgs phase of a gauge theory, since the charge approaches zero at large $R$, it annihilates {\em all} states in the theory. Hence we have
\be
|\psi'\rangle=\hat{S}|\psi\rangle\to|\psi\rangle,\,\,\,\,\,R\to\infty,
\ee
and this includes the vacuum as a special case. This again shows that there is no SSB, as there is in fact no good symmetry to spontaneously break. Instead in the Higgs phase of a gauge theory, the ``symmetry" operator has become the identity operator $\hat{S}=\hat{I}$, which is a mere redundancy. (Recall from Section \ref{NumberVacua} that at finite, but large, system size $R\gg\lamC$, the overlap between the vacuum states $|\vac\rangle$ and $|\vac'\rangle$ is exponentially close to 1, so the symmetry operator $\hat{S}$ is too.)

\section{Non-Abelian Higgs Theories}\label{NonAbelian}

In the above sections, we considered the abelian Higgs model for simplicity. However, the actual Higgs in the Standard Model is a doublet under the non-abelian gauge group $SU(2)$. Such non-abelian theories have interesting self-interactions between the spin 1 particles, meaning that we can no longer analytically construct the ground state wave-function, even in the limit in which the Higgs is taken to be heavy, as we did earlier in Section \ref{GroundStateWavefunctionals}. Nevertheless we can be confident that the central result of only a unique vacuum is still true even in this more complicated context, as we now explain.

Consider some non-abelian gauge group, such as $SU(\tilde{N})$ (or $SO(\tilde{N})$). We then have $\tilde{N}^2-1$ (or $\tilde{N}(\tilde{N}-1)/2$) spin 1 particles, which we label with an index $a$. Consistency with Lorentz symmetry \cite{Weinberg:1964ew} (or just locality \cite{Hertzberg:2017nzl}) demands that these spin 1 particles must be coupled to conserved currents. Then in the Higgs phase, these spin 1 particles acquire a mass. Suppose we again freeze the Higgs at its VEV, and that it gives each spin 1 particle the same mass $m_A$, then the equation of motion for spin 1 fields is
\be
\partial_\mu F^{a\mu\nu} = - g\,f^{abc}A_\mu^b\, F^{c\mu\nu} - m_A^2\,A^{a\nu} + J_m^{a\nu},
\label{CurrentNA}\ee
where $f^{abc}$ are the structure constants and we have  included a current $J_m^{a\nu}$ associated with matter on the right hand side. Note that if $J_m^{a\nu}$ is conserved, then we have a residual ``custodial" symmetry, which is not of importance here, as it acts trivially on the vacuum.

By integrating the time component of Eq.~(\ref{CurrentNA}) over all space, we obtain a formally conserved quantity. This is one of several color charges in the theory
\be
Q^a = \int\! d^3x\,\nabla\cdot{\bf E}^a = \oint d{\bf S}\cdot{\bf E}^a.
\label{ChargeGaussNA}\ee
Now solving for ${\bf E}^a$ from even a classical source can be complicated due to the non-linearities in Eq.~(\ref{CurrentNA}) from self-interactions. However this formula for the charge tells us that it is once again only a boundary term. Near the boundary we expect the fields to fall towards zero. Hence in this regime the non-linear terms can be ignored. This leads to the usual exponential fall off of ${\bf E}^a$ with distance, i.e., we still have $|{\bf E}^a|\propto e^{-m_A\,R}/R$ for $R\gg 1/m_A$. This ensures that each of the non-abelian charges is also zero in the Higgs phase. Hence they each generate a trivial symmetry $\hat{S}^a = \hat{I}$, which ensures that the vacuum is unique.

\section{Large-Gauge Transformations}\label{LargeGauge}

We would like to give a brief discussion of large-gauge transformations (we again use the abelian case for illustrative purposes). In this case $\alpha({\bf x})$ is non-zero at spatial infinity, but is also non constant. Following the Noether theorem, this corresponds to a family of currents $J^\mu=\partial_\nu(\alpha\, F^{\mu\nu})$. The corresponding charges are
\be
Q_\alpha = \int\! d^3x\,\nabla\!\cdot\!(\alpha({\bf x})\,{\bf E}) = \oint d{\bf A}\cdot(\alpha({\bf x})\,{\bf E}),
\ee
where we have taken the integration volume to be infinite here to ensure the charges are conserved. A family of conserved charges arises from taking $\alpha({\bf x})$ to be independent of radius, but proportional to a spherical harmonic
\be
\alpha({\bf x}) \propto Y_{lm}(\varphi,\tilde\varphi),
\ee
where $(\varphi,\tilde\varphi)$ are polar angles on the 2-sphere at spatial infinity. There is an associated set of large-gauge symmetries $\hat{S}_\alpha=\exp(i\,\hat{Q}_\alpha)$. 

In the regular phase, these charges are non-zero and act non-trivially on generic states in the theory. In particular, these charge operators do {\em not} annihilate the vacuum (except in the special case of constant $\alpha$ with $l=m=0$, which is the global symmetry.) Instead, these charges act on the vacuum and generate new vacuum states, leading to SSB. The Goldstone boson is plausibly interpreted as the massless spin 1 particle itself; the photon, or gluons above the QCD phase transition, or $W_{1,2,3}$ bosons above the electroweak scale.

In the Higgs phase, the exponential fall off of the electric field once again ensures that all these charges vanish
\be
Q_\alpha=0,
\ee
for all spherical harmonics. (The only way to try to avoid the vanishing of $Q_\alpha$ would be to make $\alpha$ grow exponentially at large radius to compensate. However this is not useful, because then one cannot justify ignoring the boundary term $\int d^3 x\,\nabla\cdot{\bf J}_\alpha$, leading to a non-conserved charge.) Hence all such corresponding ``symmetry" operators have become trivial $\hat{S}_\alpha =\hat{I}$ (i.e., redundancies) in the Higgs phase, ensuring that the vacuum is unaltered.

Hence the SSB that leads to massless spin 1 particles in the regular phase is completely removed in the Higgs phase. In this sense, the Higgs mechanism is the precise opposite of SSB (for discussion in 2+1 dimensions see Ref.~\cite{Kovner}).

\section{Defects}\label{TopologocalDefects}

An interesting feature of SSB of global symmetries is the existence of topological defects, such as domain walls in the case of a spontaneously broken $Z_2$ symmetry, or cosmic strings in the case of a spontaneously broken $U(1)$ symmetry. In the gauge case, we showed above that in fact there is a unique vacuum and no SSB, so one might wonder what is the fate of these defects. 

To address this, let us again consider the abelian Higgs model, with no fermionic sources. We search for static solutions, which readily requires $A^0=0$. Since there is in fact only a unique vacuum, we can make this manifest by operating in unitary gauge $\theta=0$, leaving the radial Higgs field $\rho$ and the massive vector ${\bf A}$ as the only degrees of freedom. The behavior of the Higgs away from its VEV will be important here, so we will no longer keep the Higgs frozen.

Let us search for axisymmetric solitonic solutions of the classical field theory, namely cosmic strings. A family of solutions in cylindrical co-ordinates $(r,\varphi,z)$ is of the form
\be
{\bf A} = \Aphi(r)\,\hat\varphi,\,\,\,\,\,\rho=\rho(r),
\ee
where $\hat\varphi$ is a unit vector in the azimuthal direction. In this ansatz, the energy (per unit length/radian) $\mathcal{E}\equiv E/(2\,\pi\,L)$ is
\bea
\mathcal{E} = \!\int_0^\infty\! dr\,r\!\left[{1\over 2}\rho'^2+V(\rho)+{1\over2}\!\left(\!\Aphi'+{\Aphi\over r}\!\right)^{\!2}\!+{1\over2}g^2\rho^2\Aphi^2\right]\!\!.  \,\,\,\,\,\,\,\,\,\,\,\,\,\,\,\label{Eeqn}
\eea

We need to introduce boundary conditions for these fields. To obtain a cosmic string centered at $r=0$, we need $\Aphi$ to be as divergent as possible as $r\to0$. In order to avoid infinite energy from the 3rd term in Eq.~(\ref{Eeqn}) we need $\Aphi$ to have the following special divergent behavior
\be
\Aphi \to {A_0\over r},\,\,\,\,\,\mbox{as}\,\,\,r\to0.
\label{Asmallr}\ee
for some constant $A_0$. Then in order to avoid infinite energy from the 4th term in Eq.~(\ref{Eeqn}) we need $\rho$ to approach zero. Lets parameterize its approach as
\be
\rho \to \rho_0\,r^n,\,\,\,\,\,\mbox{as}\,\,\,r\to0,
\label{rhosmallr}\ee
for some power $n > 0$ and prefactor $\rho_0>0$. Furthermore, we demand that the energy density falls off rapidly as $r\to\infty$ to ensure that the total energy of the cosmic string is finite. This requires both $\rho$ and $\Aphi$ to rapidly relax to the unique vacuum
\be
\rho \to v={\mu\over\sqrt{\lambda}},\,\,\,\,\,
\Aphi \to 0,\,\,\,\,\,\mbox{as}\,\,\,r\to\infty.
\ee

By varying the Hamiltonian, the corresponding classical equations for the fields $\rho$ and $\Aphi$ are the following pair of ODEs
\bea
\rho''+{1\over r}\rho' +\mu^2\,\rho & = & \lambda\,\rho^3+g^2\,\rho\,\Aphi^2,\label{rhoeqn}\\
\Aphi''+{1\over r}\Aphi'-{1\over r^2}\Aphi & = & g^2\,\rho^2\,\Aphi,\label{Aeqn}
\eea
where we have put all linear terms on the LHS and all non-linear (cubic) terms on the RHS. 

In the vicinity of $r\to0$ we insert Eqs.~(\ref{Asmallr},\,\ref{rhosmallr}) into Eq.~(\ref{rhoeqn}). We see that the only way the terms can balance is that the constant of proportionality is
\be
A_0={n\over g}.
\ee

\begin{figure}[t]
\centering
\includegraphics[width=\columnwidth]{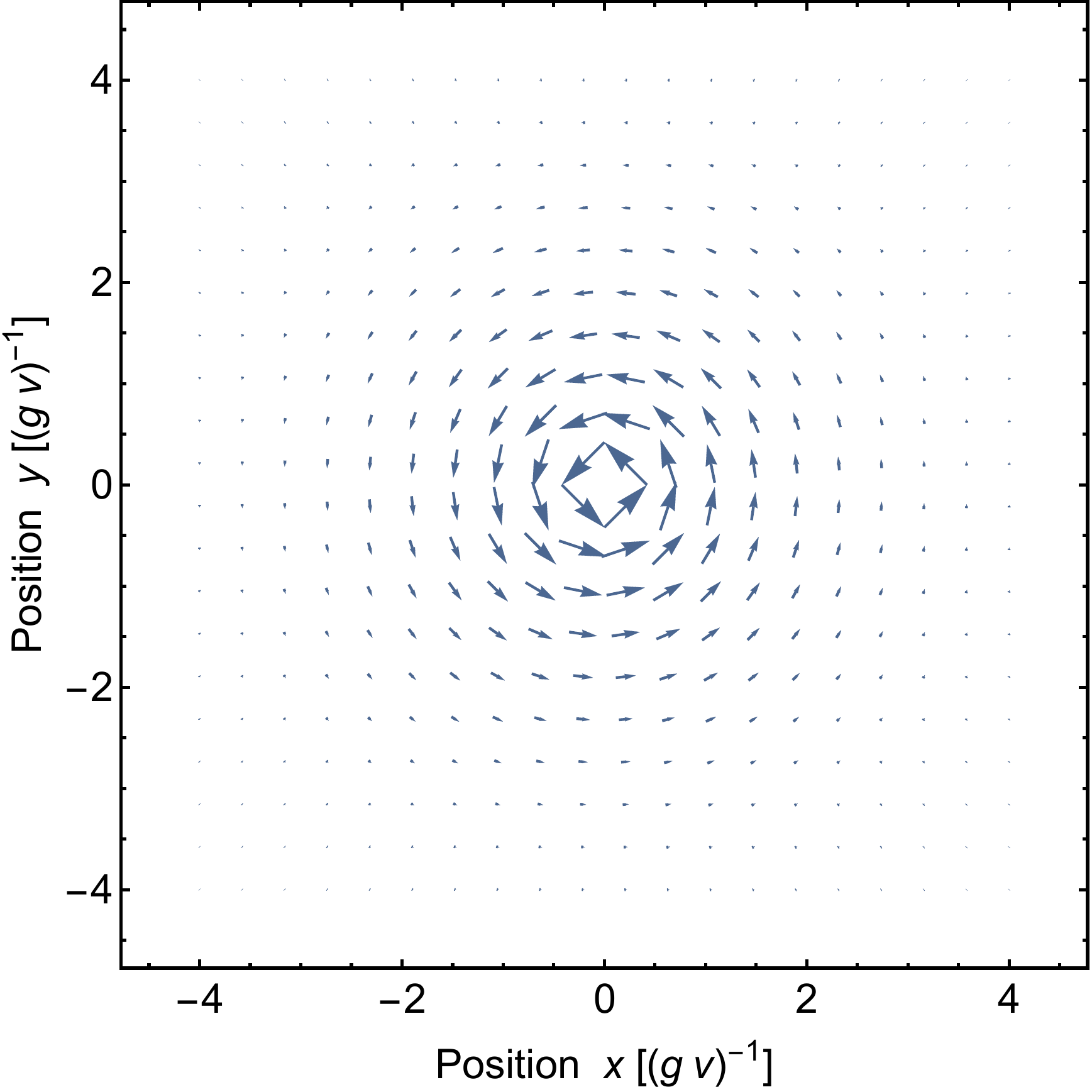}
\caption{The vector potential ${\bf A}$ (multiplied by a factor of $\rho$ for convenience) as a function of position in the $xy$-plane. The cosmic string is coming out of the page along the $z$-axis. We have chosen solution $n=1$ and coupling $\lambda=4\,g^2$ here. We are working in unitary gauge in which it is manifest that there is a unique vacuum at large radii from the core of the string. Note that the field ${\bf A}$ asymptotes to zero exponentially at large radii.}
\label{FigureCosmicString}
\end{figure}

Numerical evaluation of the above ODEs leads to the field configuration given in Fig.~\ref{FigureCosmicString}. Importantly, we note that the fields $\rho$ and $\Aphi$ asymptote to the vacuum exponentially quickly at large $r$. This is because there are only massive degrees of freedom, and so fields are naturally driven to the vacuum. This is a clear signature of having a {\em unique} vacuum. In the global case, there are a collection of vacua, and an associated massless Goldstone boson $\theta$ that slowly relaxes towards these vacua in different parts of space. However, in the gauge case, there is no Goldstone mode and so the fields approach the unique vacuum exponentially quickly. So far away from the core of the string there is no physical activity; the magnetic field and the energy density are exponentially small.

An important issue to consider is the value of the power $n$ that parameterizes this family of solutions. In this unitary gauge it appears as though any $n>0$ is a valid solution. On the other hand, there are other gauges, such as Lorenz gauge in which there is a phase field $\theta$, whose value is readily found to be $\theta=n\,\varphi$. Then by demanding the field $\cphi=\rho\,\exp(i\,\theta)/\sqrt{2}$ is well defined around a loop, one infers that $n$ must be an integer \cite{Nielsen:1973cs}. While this argument is technically correct, it gives the impression there is some physical field $\theta$ that is interpolating between different vacua at spatial infinity. However we know that when analyzed properly, there is only a unique vacuum and all physical fields are exponentially small at large radii; so this interpretation is misleading.

Instead the discreteness of $n$ can be fully understood without any appeal to multiple vacua. To do so, let us recall that the underlying theory of a solitonic solution is a quantum mechanical condensate of bosons. It is well known that consistency with the quantum mechanical particle description demands that the magnetic flux $\Phi_B$ be quantized. If we integrate the ${\bf B}$ field over a large disk whose axis is parallel to the cosmic string axis, the quantum wave-function leads to
\be
\Phi_B=\int\! d{\bf S}\cdot{\bf B}={2\,\pi\,m\over g},
\label{MagFlux1}\ee
where $m$ is an integer.

Let us now compute the magnetic flux for the above classical cosmic string solution. To do so we re-write ${\bf B}$ in terms of the vector potential ${\bf A}$ as
\be
{\bf B}=\nabla\times{\bf A} = \nabla\times\left({\bf A}-{n\over g\,r}\hat{\varphi}\right).
\ee
In the second step we have added the curl of a gradient which is zero. This is a useful trick because this subtracts off the singular piece of ${\bf A}$. With only finite quantities, we are now in a position to utilize Stoke's theorem to re-write the magnetic flux as a boundary line integral
\be
\Phi_B=\oint d{\bf l}\cdot\left({\bf A}-{n\over g\,r}\hat\varphi\right).
\label{MagFlux2}\ee
Since the vector potential relaxes to the unique vacuum $A_\varphi\to0$ at large radius, we can ignore the first term here. The second term is easily integrated to give
\be
\Phi_B = -{2\,\pi\,n\over g}.
\ee
Comparing Eqs.~(\ref{MagFlux1}) and (\ref{MagFlux2}) we learn that $n=-m$ is an integer. 

Hence we have derived that the parameter $n$ that appears in the classical field string solution must be an integer in order for the classical solution to approximate a condensate of quantum particles. Importantly, this derivation in unitary gauge does not rely upon any appeal to a classical phase field that is winding around the string, nor any reference to multiple vacua. Instead there is manifestly only one unique vacuum and the cosmic string is simply a condensate of particles whose behavior is restricted by the quantization of angular momentum.

In a cosmological context, defects can emerge in a phase transition from the regular phase to the Higgs phase. They can comprise a significant component of the energy density of the universe. However, as the universe expands their average energy densities $u$ evolve as
\be
u(t) \propto {1\over a(t)^{3(1+w)}},
\ee
where $a(t)$ is the scale factor, with $w=-2/3$ for domain walls, $w=-1/3$ for cosmic strings, $w=0$ for monopoles, and $w>0$ for textures. In all cases, since $w>-1$, the average energy density $u \to 0$ at late times, and the system evolves toward the unique vacuum. So in much the same way that matter and radiation at late times are understood as excitations above the vacuum, defects in a gauged Higgs model, such as cosmic strings, should be understood as excitations above the (unique) vacuum too.

\section{Path Integral Measure}\label{PathIntegralMeasure}

Having shown that there is a unique vacuum in a gauged Higgs model, it is important to ascertain the correct measure on the quantum path integral for the radial Higgs mode $\rho$. Let us again illustrate the issue within the context of an abelian Higgs model, before generalizing to the non-abelian case. 

On the one hand, if we go to Coulomb gauge $A^L=0$, we could use the cartesian variables for a complex Higgs field $\cphi$ and $\cphi^*$, or switch to polar variables $\rho$ and $\theta$. It would directly suggest that the partition function have the form
\be
Z=\int\!\mathcal{D}\!\cphi\,\mathcal{D}\cphi^*\mathcal{D}A^T\,e^{i\,S_C}= \int\!\mathcal{D}\rho\,\rho\,\mathcal{D}\theta\,\mathcal{D}A^T\,e^{i\,S_C},
\label{Measure1}\ee
where $S_C$ is the Coulomb gauge fixed action. The integration $\int\!\mathcal{D}\rho\,\rho$ is a ``volume" measure in the 2-dimensional scalar field space. On the other hand, we know this Coulomb gauge picture can be highly misleading. In particular, it gives the illusion of many different vacua labelled by different values of $\theta$, when in fact we earlier proved that the vacuum is unique. A way to make this manifest is to switch to unitary gauge in which $\theta=0$. In this case the field theory is described by a single scalar field $\rho$ (along with a massive spin 1 vector). From this point of view it seems reasonable to suppose that the partition function is
\be
Z=\int\!\mathcal{D}\rho\,\mathcal{D}A^T\mathcal{D}A^L\,e^{i\,S_U},
\label{Measure2}\ee
where $S_U$ is the unitary gauge fixed action. Here there are no additional factors of $\rho$; the integration $\int\!\mathcal{D}\rho$ is a ``line" measure in an effective 1-dimensional scalar field space. The additional factor of $\rho$ that appears in the volume measure in Eq.~(\ref{Measure1}) appears to imply that one is counting many different vacua labelled by different $\theta$. While the line measure in Eq.~(\ref{Measure2}) arises from the idea that this volume measure fails to take into account the redundancy of the gauge theory in the $\theta$ direction, collapsing the effective scalar field space to a line. It appears unclear which is the correct measure.

Here we derive the correct answer by demanding consistency of the quantum theory. Let us continue to work in unitary gauge $\theta=0$. We expand the Higgs around its VEV as $\rho=v+h$, with (bare) Lagrangian density
\bea
\mathcal{L}_0 = -\dfrac{1}{4}F_{\mu\,\nu}F^{\mu\,\nu} +{1\over2}g^2\,v^2\,A_\mu^2 + {1\over2}(\partial_{\mu}h)^2 - \lambda\,v^2\,h^2\,\nonumber\\
+ g^2v\,h\,A_\mu^2+ {1\over2}g^2\,h^2\,A_{\mu}^2 - \lambda\,v\,h^3-{1\over4}\lambda\,h^4,
\label{LagUnitary}
\eea
where the quadratic terms are written on the first line and the interaction terms are written on the second line.

To make progress we work perturbatively. Importantly, a massive 1 particle has a peculiar form for its propagator. In momentum space it is
\be
P_{\mu\,\nu}(k)  = -i\dfrac{\eta_{\mu\,\nu}-{k_\mu k_\nu\over g^2\,v^2}}{k^2 - g^2\,v^2+i\,\epsilon},
\ee
(with $k^2=k_\mu k^\mu$ here). What is interesting is that the longitudinal mode contributes the additional term to the numerator. This implies the propagator is not decreasing at high momenta, but is instead constant
\be
\eta^{\mu\nu}P_{\mu\nu}(k) \to P_\infty \equiv {i\over g^2\,v^2},\,\,\,\,\mbox{as}\,\,\,k\to\infty.
\ee
This leads to strong UV divergences in loop integrals, which we now compute.

There are 3-point and 4-point interactions between the Higgs and the spin 1 particle given by the first two terms on the second line of Eq.~(\ref{LagUnitary}). The Feynman rules are
\bea
\begin{tikzpicture}
\begin{feynman}
\vertex (a);
\vertex [right=1cm of a] (b);
\vertex [above right=1cm of b] (c);
\vertex [below right=1cm of b] (d);
\vertex [right=1cm of b] (e) {\( =  2\,i\,g^2\,v\,\eta_{\mu\,\nu},\)};
\diagram*{
(a) -- [solid] (b),
(b) -- [photon] (c),
(b) -- [photon] (d),
};
\end{feynman}
\end{tikzpicture}
\,\,\,
\label{eq:FeynmanRule1}
\eea
\bea
\begin{tikzpicture}
\begin{feynman}
\vertex (g);
\vertex [above left=1cm of g] (h);
\vertex [below left=1cm of g] (i);
\vertex [above right=1cm of g] (j);
\vertex [below right=1cm of g] (k);
\vertex [right=1cm of g] (f) {\( = 2\,i\,g^2\,\eta_{\mu\,\nu}.\)}; 
\diagram*{
(h) -- [solid] (g),
(i)  -- [solid] (g),
(g) -- [photon] (j),
(g) -- [photon] (k),
};
\end{feynman}
\end{tikzpicture}
\label{eq:FeynmanRule2}
\eea
These interactions generate non-trivial correlation functions involving $N$ external Higgs particles (there are also corrections from Higgs self-interactions, but such effects will not contribute to the leading UV divergences).

Let us focus here on divergences that arise at one-loop. The 1-point amplitude for the Higgs has the contribution
\bea
&&\hspace{-0.1cm}
\begin{tikzpicture}
\begin{feynman}
\vertex (a);
\vertex [right=1cm of a] (b);
\vertex [right=0.75cm of b] (c);
\vertex [right=0.2cm of c] (d); 
\vertex [left=1.2cm of b] (x) {\( i\,M_1=\)};
\diagram*{
(a) -- [solid] (b),
(b) -- [photon, out=270, in=270, loop, min distance=0.5cm] (c),
(c) -- [photon, out=90, in=90, loop, min distance=0.5cm] (b),
};
\end{feynman}
\end{tikzpicture} \nonumber\\
&& 
= \dfrac{1}{2}(2\,i\,g^2\,v)\int\!\dfrac{d^4k}{(2\pi)^4}\,\eta^{\mu\,\nu}\,P_{\mu\,\nu}(k)\xrightarrow{UV} - \dfrac{\Lambda^4}{v}.
\label{1point}\eea
To regulate the integral we have introduced a UV cutoff $\Lambda$, defined by 
\be
\int \! {d^4k\over(2\pi)^4} = \Lambda^4.
\ee
There are sub-leading terms to the integral in Eq.~(\ref{1point}) that are not of direct importance here. 

Similarly, it can be shown that all the higher $N$-point functions have a leading quartic divergence too. For the 2-point amplitude $i\,M_2$ we find two contributions
\bea
&&\hspace{-0.1cm}
\begin{tikzpicture}
\begin{feynman}
\vertex (a);
\vertex [right=1cm of a] (b);
\vertex [right=1cm of b] (c);
\vertex [above=0.75cm of b] (d);
\vertex [right=1.3cm of b] (e) {\(+\)};
\vertex [right=0.6cm of e] (f);
\vertex [right=0.8cm of f] (g);
\vertex [right=0.75cm of g] (h);
\vertex [right=0.8cm of h] (i);
\vertex [left=1.2cm of b] (x) {\( i\,M_2=\)};
\diagram*{
(a) -- [solid] (b),
(b) -- [solid] (c),
(b) -- [photon, out=45, in=315, loop, min distance=0.5cm] (d),
(d) -- [photon, out=225, in=135, loop, min distance=0.5cm] (b),
(f) -- [solid] (g),
(g) -- [photon, out=270, in=270, loop, min distance=0.5cm] (h),
(h) -- [photon, out=90, in=90, loop, min distance=0.5cm] (g),
(h) -- [solid] (i),
};
\end{feynman}
\end{tikzpicture} \nonumber\\
&&\xrightarrow{UV}  \dfrac{1}{2}(2\,i\,g^2)\,\Lambda^4P_\infty + \dfrac{1}{2}(2\,i\,v\,g^2)^2\Lambda^4P_\infty^2 = \dfrac{\Lambda^4}{v^2}.\,\,\,\,\,\,\,\,\,\,\,
\label{2point}\eea
The 3-point amplitude $i\,M_3$ also has two contributions
\bea
&&\hspace{-0.1cm}
\begin{tikzpicture}
\begin{feynman}
\vertex (a);
\vertex [right=1cm of a] (b);
\vertex [right=0.75cm of b] (c);
\vertex [above right=1cm of c] (d);
\vertex [below right=1cm of c] (e);
\vertex [right=1cm of c] (f) {\(+\)};
\vertex [right=0.6cm of f] (g);
\vertex [right=1cm of g] (h);
\vertex [above right=1cm of h] (i);
\vertex [below right=1cm of h] (j);
\vertex [right=1cm of i] (k);
\vertex [right=1cm of j] (l);
\vertex [left=1.2cm of b] (x) {\( i\,M_3=\)};
\diagram*{
(a) -- [solid] (b),
(b) -- [photon, out=270, in=270, loop, min distance=0.5cm] (c),
(c) -- [photon, out=90, in=90, loop, min distance=0.5cm] (b),
(c) -- [solid] (d),
(c) -- [solid] (e),
(g) -- [solid] (h),
(h) -- [photon] (i),
(i) -- [photon] (j),
(j) -- [photon] (h),
(i) -- [solid] (k),
(j) -- [solid] (l),
};
\end{feynman}
\end{tikzpicture} \nonumber\\
&& \xrightarrow{UV} \dfrac{3}{2}(2\,i\,g^2\,v)(2\,i\,g^2)\,\Lambda^4P_\infty^2
+ (2\,i\,v\,g^2)^3\Lambda^4P_\infty^3 = -2\,\dfrac{\Lambda^4}{v^3}.\,\,\,\,\,\,\,\,\,\,\,\,\,\,
\label{3point}\eea
With similar behavior for all higher order functions. The contribution to the amplitude $i\,M_N$ for the general $N$-point function is found to be
\be
i\,M_N \xrightarrow{UV} (-1)^N(N-1)! \,{\Lambda^4\over v^{N}}.
\ee

In order to cancel each of these quartic divergences we need to add a tower of counter terms to the Lagrangian density. These must be powers of $h^N$ with appropriate coefficients to achieve cancellation, namely
\be
\mathcal{L}_{ct} = -i\,{\Lambda^4\over v}h+i{\Lambda^4\over 2\,v^2}h^2-i{\Lambda^4\over 3\,v^3}h^3+\ldots+i(-1)^N\!{\Lambda^4\over N\,v^N}h^N+\ldots
\ee
Note that this series sums to infinity for $h<-v$ (while it is absolutely convergent for $|h|<v$, conditionally convergent for $h=v$, and can be regulated to a finite answer for $h> v$). So the unitary gauge theory is telling us that $\rho=v+h>0$ for finiteness. In this domain we can resum this tower of terms to a logarithm
\be
\mathcal{L}_{ct}=-i\,\Lambda^4\,\ln(v+h)=-i\,\Lambda^4\,\ln\rho,
\ee
(up to an irrelevant constant). 

Now the physical meaning of the cutoff $\Lambda$ in physical space is that we are effectively defining the field theory on a space-time lattice with spacing $1/\Lambda$. So the corresponding correction to the unitary gauge action $S_U=S_0+S_{ct}$ is 
\be
S_{ct} = \int d^4x\,\mathcal{L}_{ct} = -i\sum_{lattice} \ln\rho.
\ee
By exponentiating and inserting this counter term into the unitary gauge partition function Eq.~(\ref{Measure2}), we obtain
\be
Z=\int\mathcal{D}\rho\,\rho\,\mathcal{D}A^T\mathcal{D}A^L\,e^{i\,S_0}.
\label{Measure3}\ee
Hence we obtain a factor of $\rho$ in the path integral measure in order to ensure renormalizability. We have also checked that this is maintained at two-loop order. This implies that indeed the volume measure for the Higgs is correct.

This is not too surprising given that earlier in Section \ref{Spectrum} we explained that there is a one-to-one correspondence between the states in the global case and the gauge case. It is obvious that in the global case we need a volume measure since we evidently have a pair of cartesian fields, without any gauge ambiguity to contend with. So one should anticipate there must also be a volume measure in the gauge case, it is simply that the degeneracy of the vacuum has been lifted in this case.

We note that this result has a simple generalization beyond the abelian $U(1)$ Higgs model. If we have an $SU(\tilde{N})$ (or $SO(\tilde{N})$) model, with the Higgs transforming in the fundamental representation of $SU(\tilde{N})$ (or $SO(\tilde{N})$), then the Higgs multiplet is comprised of $N=2\,\tilde{N}$ (or $N=\tilde{N}$) real scalars. Of these we can identify one as the radial Higgs field $\rho$, along with the remaining $N-1$ would-be Goldstones that may be removed in unitary gauge or maintained in other gauges such as Coulomb. In any case, the path integral measure $d\mathcal{M}_{\rho}$ for the radial Higgs field $\rho$ is found to be
\be
d\mathcal{M}_{\rho} \propto \mathcal{D}\rho\,\rho^{N-1},
\label{Measure4}\ee
in order for the theory to be renormalizable.

\section{Application to Cosmology}\label{ApplicationtoCosmology}

The measure in Eq.~(\ref{Measure4}) has potentially interesting ramifications for cosmology. Consider the Higgs of the Standard Model, which is a complex doublet, transforming in the fundamental representation of $SU(2)_L\times U(1)_Y$. Suppose the Standard Model is valid up to very high energies. In this case the classical potential for the Higgs $V(\rho)$ (Eq.~(\ref{HiggsPotential})) undergoes significant corrections from renormalization. At one-loop the beta-function for the Higgs self-coupling $\lambda$ in the Standard Model is well known to be (e.g., see \cite{Ford:1992pn})
\be
\beta_\lambda = {1\over(4\pi)^2}\!\left[24 \lambda ^2-6\sum_fy_f^4+\frac{3}{8} \left(2 g^4+\left(g^2+g'^2\right)^2\right)\right]\!,
\ee
with $\beta_\lambda \equiv d\lambda/d\ln E$. Here $y_f$ are the fermion Yukawa couplings, $g$ is the $SU(2)_L$ coupling, and $g'$ is the $U(1)_Y$ coupling. Due to the large top quark Yukawa coupling $y_t\sim 1$, and using measured values for $\lambda,\,g,\,g'$, one finds $\beta_\lambda<0$. So even though $\lambda\sim 0.1>0$ below the electroweak scale (giving $m_h=\sqrt{2\,\lambda}\,v\approx 125$\,GeV) it becomes negative at much higher energies due to this logarithmic running. This means the renormalized potential $V_{ren}(\rho)$ turns over and goes negative, rendering our vacuum meta-stable. Precise renormalization group computations at high loop order reveal that the turn over is around $E^*\sim 10^{11}$\,GeV, or so, depending on the precise value of parameters. This is the zero temperature result. There are corrections at very high temperatures, which may be relevant to the early universe. But within the framework of the Standard Model, the finite temperature effective potential still has a turn over, albeit at even higher values, and so this is still important.

An important question then is: what is the probability that in the very early universe the Higgs began on the favorable side of the potential $\rho<E^*$ in order to relax to our electroweak vacuum? If one thought the measure is a line measure $d\mathcal{M}_\rho \propto \mathcal{D}\rho$, since there is gauge redundancy, one would infer the probability is $p_1$, where $p_1$ is the probability of one scalar field having an initial value $<10^{11}$\,GeV (or the corresponding high temperature turn over value). This may be a small probability ($p_1\ll 1$) during the very early universe (perhaps near the Planck era) where typical field values may have been Planckian $\sim 10^{18}$\,GeV. On the other hand, since we proved that the correct measure is actually the volume measure $d\mathcal{M}_{\rho} \propto \mathcal{D}\rho\,\rho^{N-1}$, the correct probability for the Higgs scales roughly as
\be
p_\rho\sim p_1^N = e^{-N\,\ln(1/p_1)},
\ee
where $N=4$ for the Standard Model Higgs doublet. Hence the probability for the Higgs to have begun on the favorable side is exponentially small in $N(=4)$. This rigorously justifies the probability estimates that were used in Ref.~\cite{Hertzberg:2012zc} which also addressed this issue.

\section{Outlook}\label{Discussion}

We have rigorously shown that the ground state of a gauged Higgs theory is unique. To do so, we showed that the overlap between vacuum states rapidly approaches one at finite gauge coupling in the limit in which the volume of space is taken to infinity, leaving a unique vacuum. It is only in the limit in which the gauge coupling is sent to zero, does one have a small overlap and obtain multiple vacua. This counting of ground states is the precise way of addressing the question of whether the Higgs mechanism leads to spontaneous symmetry breaking, which we have answered in the negative. We showed that this is reinforced by the energy spectrum, the structure of the canonically normalized Lagrangian, the triviality of conserved charges in the Higgs phase. We noted that it is consistent with the existence of defects.

However, despite the presence of only a unique vacuum, the correct measure on the quantum path integral is $d\mathcal{M}_{\rho} \propto \mathcal{D}\rho\,\rho^{N-1}$ in accord with having $N-1$ would-be Goldstones. This is a gauge independent statement, regardless of whether they are described as scalars in Coulomb gauge, or described more directly as longitudinal modes of massive spin 1 particles in unitary gauge. This measure on states has potential ramifications for cosmology.

Important investigations would be to determine the probability $p_1$ for the Higgs to begin on the favorable side of the potential in the context of inflation, both low scale and high scale, and eternal inflation, or other scenarios for the early universe. A much more detailed treatment of probabilities in the early universe will be addressed in a forthcoming paper \cite{HertzbergJainInflation}. One should also consider this issue in the context of grand unification, where one has $N=10$ in $SU(5)$ and $SO(10)$ models, as well as string models with large gauge groups, in which this exponential change in probability is even more extreme. We leave these topics for future work.

\section*{Acknowledgments}

We would like to thank  Larry Ford, Tony Gherghetta, Gary Goldstein, Alex Kovner, Vijay Kumar, Mehrdad Mirbabayi, Ken Olum, McCullen Sandora, Alex Vilenkin, and Masaki Yamada for useful discussions. MPH is supported in part by National Science Foundation grant PHY-1720332.

\end{document}